\documentclass{article}
\usepackage{arxiv}
\usepackage{graphicx}
\usepackage{subcaption}
\usepackage{xcolor}
\usepackage{hyperref}
\usepackage{url}
\usepackage[usenames,dvipsnames]{pstricks}
\usepackage{epsfig}
\usepackage{pst-grad} % For gradients
\usepackage{pst-plot} % For axes
\usepackage{amsmath,epsfig}
\usepackage{amssymb}
\usepackage{psfrag}
\usepackage{color}
\usepackage{pstricks}
\usepackage{cite}
\usepackage{algorithm}
\usepackage{epsfig}
\usepackage{amsthm}
\usepackage{amssymb}
\usepackage{psfrag}
\usepackage{pstricks}
\usepackage{float}
\usepackage{stfloats}
\usepackage{wrapfig}
\usepackage{amsfonts}
\usepackage{amssymb}
\usepackage{blindtext}
\usepackage{array}
\usepackage{arydshln}
\usepackage{pstricks,pst-plot}
\usepackage{pst-bezier}
\usepackage{pst-math}
\usepackage{algorithm,algcompatible}
\usepackage{float}
\usepackage[mathscr]{eucal}
\usepackage{enumerate}
\usepackage{hyperref}
\usepackage{multirow}
\usepackage{psfrag}
\usepackage{pstool}
\usepackage{tikz}

\usepackage{svg} 
\graphicspath{ {./images/} } \usepackage{graphicx}
\usepackage{amsmath}
\usepackage{algorithm,algcompatible,amsmath} \usepackage{booktabs} \usetikzlibrary{arrows.meta}

\def\argmax{{\rm arg}\max}

\def\b0{{\bf 0}}

\title{Scaling LLM Inference via Parallelism: Throughput-Latency Trade-off }

\author{
 \large Javad Mirzaei\thanks{Corresponding author.}\\
   Office of CTO\\
  Dell Technologies Inc. \\
  {Canada} \\
  \texttt{javad.mirzaei@dell.com} 
   \And
  \large Jeebak Mitra \\
  Office of CTO\\
  Dell Technologies Inc. \\
  {Canada} \\
  \texttt{jeebak.mitra@dell.com}
}
\begin{document}
\maketitle
\begin{abstract}
Large Language Model (LLM) serving systems must balance latency and throughput while efficiently utilizing distributed GPU resources. This challenge arises because inference consists of two fundamentally different phases: prefill, which determines time-to-first-token (TTFT), and decode, which determines inter-token latency (ITL). The use of tensor parallelism (TP), pipeline parallelism (PP), and hybrid configurations exposes different trade-offs between these latency dimensions, making the optimal parallelism configuration highly dependent on workload characteristics, latency requirements, and context length. Moreover, despite the widespread adoption of distributed inference, there is still limited understanding of how parallelism configurations should be selected under varying workload conditions and service-level objectives (SLO). In this paper, we develop a systematic framework for characterizing LLM inference parallelism via throughput-latency trade-offs, SLO-aware serving, and long-context scalability. We introduce the notion of \emph{parallelism operating regions} by constructing TTFT-ITL feasible sets and Pareto frontiers across different workload compositions, thereby revealing a systematic transition from TP-dominant to PP-dominant configurations as workloads become increasingly prefill-heavy. We further formulate \emph{SLO-aware parallelism selection} using goodput, showing that stringent TTFT requirements favor PP-dominated configurations, whereas relaxed latency constraints progressively shift the optimality toward TP-dominated configurations with higher serving capacity. Finally, we demonstrate that increasing context length fundamentally changes the relative advantages of parallelism configurations, with PP-dominant configurations exhibiting more favorable scalability and SLO attainment in the long-context regime. 
\end{abstract}

\section{Introduction}
Large language models (LLMs) have become a foundational component of modern AI applications, powering conversational assistants, code-generation systems, retrieval-augmented generation workflows, and increasingly complex agentic pipelines. As these applications move from offline processing to interactive services, inference efficiency has become a primary serving challenge. Modern LLM serving systems must simultaneously satisfy stringent latency requirements, sustain high request throughput, and efficiently utilize expensive GPU resources. Achieving these objectives is particularly difficult because autoregressive inference consists of two fundamentally different phases: prefill, which processes the input context in parallel, and decode, which generates output tokens autoregressively. These phases exhibit distinct computational, memory, and communication characteristics, which creates an inherent trade-off between time-to-first-token (TTFT) and inter-token latency (ITL) \cite{agrawal2024taming, zhong2024distserve}. Consequently, evaluating serving systems using only a single latency or throughput metric is often insufficient to characterize real-world performance. Instead, systems increasingly consider service-level objectives (SLO) and goodput\footnote{Goodput is the rate at which requests are completed while satisfying the specified latency requirement.} as serving objectives \cite{wang2025prefilldecodeaggregationdisaggregationunifying, chen2025slosserveoptimizedservingmultislo}.

The challenge is further compounded by the diversity of LLM workloads. Requests can vary substantially in input length, output length, concurrency, and latency requirements, causing different phases of the inference pipeline to become dominant under different operating conditions \cite{ruan2025dynaserveunifiedelasticexecution, wang2026tamingrequestimbalancesloaware}. There is no single parallelization configuration that is universally optimal across serving scenarios. Tensor parallelism (TP), pipeline parallelism (PP), and their hybrid configurations impose different computation, synchronization, communication, and memory trade-offs, and their efficacy depends strongly on model size, sequence length, hardware topology, and workload composition \cite{amer2026distributedhybridparallelismlarge, xu2025characterizingcommunicationpatternsdistributed,guo2025gllmglobalbalancedpipeline}. As a result, the use of parallelism should not be viewed as a static universal deployment decision; rather, it is a workload-dependent operating point that balances competing latency and throughput objectives. Dynamic reconfiguration and parallelism-switching approaches have therefore been proposed to adapt execution strategies as workload conditions change \cite{hidayetoglu2026shiftparallelismlowlatencyhighthroughput,yuan2026remplowdowntimeruntimemodelparallelism, gao2026flyingservingontheflyparallelis}.

A second challenge stems from the SLO constraints encountered in production deployments. In production environments, maximizing raw throughput is often insufficient because only requests that satisfy their latency requirements constitute useful service \cite{zhong2024distserve, adnan2026cascadeexploitingsloawarelatency}. Moreover, the relative importance of TTFT and throughput depends on application requirements. For example, interactive applications generally demand fast responsiveness, whereas batch or offline workloads may prioritize overall serving capacity and cost efficiency \cite{srivatsa2026nitsumservingtieredllm, chen2025slosserveoptimizedservingmultislo}. The problem becomes even more pronounced for long-context inference, where increasing context length increases prefill computation, communication overhead, key-value (KV) cache memory consumption, and scheduling complexity \cite{shi2026vppvirtualpipelineparallelism, guo2026adaptivecontextparallelismproduction}. Under these conditions, the preferred parallelism and scheduling strategy can change dramatically as sequence lengths, workload heterogeneity, and latency requirements evolve, thereby making static parallelism configuration choices inefficient \cite{song2026deadlineawareadaptiveprefillchunking, hidayetoglu2026shiftparallelismlowlatencyhighthroughput}.

In this paper, we develop a systematic framework to understand and select parallelism configuration for LLM inference through three complementary perspectives: latency trade-offs, SLO-aware serving, and long-context scalability. First, we introduce the notion of a \emph{parallelism operating region} by constructing TTFT-ITL feasible sets and Pareto frontiers across different workload compositions, revealing a systematic transition from heavy TP-dominant to PP-dominant configurations as workloads move from decode-heavy to prefill-heavy regimes. Second, we formulate \emph{SLO-aware parallelism selection}, which enables parallelism configurations to be evaluated according to both serving capacity and latency compliance. We show that the optimal configuration shifts from PP-dominant to TP-dominant strategies as TTFT requirements are relaxed. Finally, we extend the analysis to long-context serving and show that sequence-length growth fundamentally changes the relative advantages of different parallelism strategies, increasingly favoring PP-dominant configurations under stringent latency constraints. Together, these results provide a unified view of how workload characteristics, latency requirements, and context length jointly determine the optimal operating region for LLM inference parallelism, thereby offering practical guidance for deploying and adapting serving systems across diverse application scenarios.
\subsection{Related Works}
\textbf{Latency-oriented LLM serving}:
Recent works on LLM serving have shown that throughput alone is insufficient to characterize system performance, mainly because TTFT and ITL depend on distinct phases of inference and exhibit different sensitivities to system design choices. During prefill, computation scales with input sequence length, whereas decoding proceeds iteratively while repeatedly accessing a growing KV cache. Consequently, configurations that maximize token throughput may still deliver poor interactive performance. Recently, attention has shifted toward latency-aware metrics. Sarathi-Serve \cite{agrawal2024taming} examines the trade-off among batching, throughput, and streaming latency, while DistServe \cite{zhong2024distserve} introduces goodput as the throughput of requests that satisfy latency requirements. This perspective is further reinforced by holistic evaluation frameworks in \cite{agrawal2024etalonholisticperformanceevaluation,usami2026prefilldecodeawareevaluationllminference}, which show that prompt length, output length, request arrival patterns, and hardware topology jointly determine performance. As a result, modern LLM serving systems are increasingly evaluated in terms of their ability to satisfy TTFT and ITL constraints rather than optimizing a single latency or throughput metric. 

The separation between prefill and decode phases has also motivated architectural innovations that explicitly address resource utilization during inference. When long prefill computations and latency-sensitive decoding requests share the same resources, contention can simultaneously degrade TTFT and ITL. To mitigate this effect, Nexus \cite{shi2025nexusproactiveintragpudisaggregationprefill} explores proactive prefill-decode disaggregation and chunked-prefill scheduling, while \cite{wang2025prefilldecodeaggregationdisaggregationunifying} studies the conditions where aggregation or disaggregation yields superior goodput. These observations suggest that the efficacy of any parallelism configuration depends not only on its computational efficiency but also on how resources are allocated between prefill and decode workloads. Consequently, the choice of a parallelism configuration is closely coupled with broader serving-system architectural decisions. 

To scale increasingly large models and long-context workloads, existing systems employ combinations of data parallelism (DP), TP, and PP. Hybrid approaches are often necessary because model weights, activations, KV caches, computation, and communication scale differently across hardware platforms and workloads \cite{amer2026distributedhybridparallelismlarge}. DP provides high request-level concurrency and benefits from efficient KV-cache management techniques such as PagedAttention \cite{kwon2023efficientmemorymanagementlarge}, while TP enables model execution beyond the memory limits of a single accelerator and has motivated adaptive mechanisms for workload-dependent reconfiguration \cite{srivatsa2026nitsumservingtieredllm,yuan2026remplowdowntimeruntimemodelparallelism}. PP further expands model capacity but introduces additional pipeline scheduling and load-balancing challenges, particularly for long-context workloads \cite{shi2026vppvirtualpipelineparallelism}. More recently, dynamic approaches such as on-the-fly parallelism switching \cite{gao2026flyingservingontheflyparallelis} have emphasized that no single parallelism configuration is universally optimal. Instead, the appropriate configuration depends on workload characteristics, latency objectives, and changing demand conditions.

\textbf{SLO-aware serving}:
Moving from throughput-centric evaluation toward TTFT and/or ITL constrained serving, recent work has increasingly adopted goodput as the primary optimization objective. Unlike raw throughput, goodput measures only the requests that satisfy their latency requirements, thereby directly reflecting user-perceived quality of service. DistServe \cite{zhong2024distserve} demonstrates how architectural decisions such as prefill-decode disaggregation can significantly improve goodput under latency constraints, while Cascade \cite{adnan2026cascadeexploitingsloawarelatency} extends this perspective through SLO-aware resource allocation for heterogeneous workloads. Together, these studies emphasize that maximizing token processing rates alone is insufficient, and effective serving systems must balance resource utilization against the ability to consistently meet latency targets across diverse request classes.

A recurring theme in the literature is the close interaction between parallelism configuration and request scheduling. Parallelism techniques determine the available compute capacity, communication overhead, and memory footprint, whereas scheduling policies decide how these resources are allocated to incoming requests. Prior work has shown that scheduling decisions strongly influence latency outcomes, especially in the presence of long input prompts and mixed workloads. To address this challenge, Sarathi-Serve \cite{agrawal2024taming} introduced chunked prefill to prevent large prompts from monopolizing execution and to enable better interleaved decode operations, thereby improving responsiveness without sacrificing throughput. Nexus further advances this idea by exploring the scheduling granularity of prefill workloads in shared-resource environments \cite{shi2025nexusproactiveintragpudisaggregationprefill}. These works collectively suggest that achieving high goodput requires coordinated optimization of both system configuration and runtime scheduling rather than addressing either problems in isolation. 

The challenge becomes significantly more pronounced in multi-SLO and heterogeneous serving environments, where requests exhibit different latency requirements, priorities, and resource demands. GoodServe  \cite{du2026goodservehighgoodputservingagentic} investigates goodput-aware serving for agentic inference workloads spanning heterogeneous resources, while Nitsum \cite{srivatsa2026nitsumservingtieredllm}, Niyama \cite{goel2025niyamabreakingsilos}, and SLOs-Serve \cite{chen2025slosserveoptimizedservingmultislo} explore adaptive parallelism, QoS-aware scheduling, and dynamic request routing to meet differentiated service objectives. Infinite-LLM \cite{lin2024infinitellmefficientllmservice} addresses the challenge of highly variable context lengths through attention-layer disaggregation. At the same time, uncertainty in output length introduces additional complexity because future KV-cache consumption and decode time are unknown at admission. Jenga \cite{zhang2025jengaeffectivememorymanagemen} addresses memory efficiency through a heterogeneity-aware allocation framework, while \cite{cheng2026robustkvcachemanagement} highlights the trade-off between conservative memory reservation and aggressive admission policies under uncertain decode demand. These studies point toward a control framework where long-term resource allocation, parallelism selection, memory management, and dynamic scheduling decisions must be jointly optimized to maximize goodput while maintaining SLO compliance.

\textbf{Serving for long context}:
As context lengths continue to grow, traditional model-parallel techniques alone become insufficient to sustain efficient LLM serving. Long-context workloads increase attention computation during prefill, expand KV-cache requirements, and impose higher memory-access costs during decoding, thereby affecting TTFT and ITL through different mechanisms. TTFT is primarily influenced by the cost of processing increasingly large input prompts, while ITL degrades as each decoding step repeatedly accesses a larger context state. Medha \cite{agrawal2025medha} highlights this challenge by targeting inference at multi-million-token context lengths, demonstrating that sequence state and KV-cache management can become the dominant scalability bottlenecks. To address this issue, recent studies have focused on context parallelism (CP), which distributes context tokens and associated state across multiple accelerators \cite{jiang2025dcpaddressinginputdynamism, niu2026dhpefficientscalingmllm, wang2026flashcploadbalancedcommunicationefficientcontext, zhao2026unleashingscalablecontextparallelism, guo2026adaptivecontextparallelismproduction, ghadia2026untiedulyssesmemoryefficientcontext, zhang2025pecschedpreemptiveefficientcluster}. Adaptive CP techniques have been explored for production environments \cite{guo2026adaptivecontextparallelismproduction}, while Untied Ulysses \cite{ghadia2026untiedulyssesmemoryefficientcontext} improves memory efficiency through headwise chunking and PecSched \cite{zhang2025pecschedpreemptiveefficientcluster} reduces communication overhead for sequence-parallel execution. Collectively, these studies suggest that CP is an effective mechanism when sequence length and KV-cache growth dominate resource consumption, although its communication costs can outweigh its benefits for shorter-context workloads.

Long-context serving also strengthens the interaction between scheduling policies, pipeline execution, and attention mechanisms. While chunked prefill has emerged as an effective mechanism for reducing interference and preserving decode responsiveness, the computational cost of successive chunks is not uniform because later chunks attend to progressively longer prefixes. The work in  \cite{shi2026vppvirtualpipelineparallelism} addresses this challenge through virtual pipeline parallelism (VPP) that better balances chunked-prefill execution under long-context workloads, thereby linking pipeline scheduling directly to the TTFT-ITL trade-off discussed earlier. Beyond dense-attention models, recent work has further shown that the choice of attention mechanism can significantly influence parallelism configuration efficacy. SparseServe \cite{zhou2025sparseserveunlockingparallelismdynamic} demonstrates that dynamic sparse-attention mechanisms introduce irregular computation and communication patterns that require specialized scheduling and distribution policies to fully exploit available parallelism. Together, these works indicate that efficient long-context serving requires a holistic approach that jointly considers context parallelism, pipeline scheduling, memory management, and attention algorithms, extending the broader theme that optimal serving performance emerges from coordinated management of both system architecture and runtime execution policies.

\section{Parallelism Operating Region} \label{sec:latency_pareto}
During inference, a request goes through prefill and decode phases, each with fundamentally different execution characteristics. In particular, during prefill, the input is processed in parallel, while during decode, output tokens are generated autoregressively. A key challenge is that these phases are not computationally homogeneous. When  multiple GPUs are used, inter-GPU communication becomes an additional challenge along with computation overheads \cite{mirzaei2026characterizingparallelismstrategiesllm}. To illustrate this effect, we represent a parallelism configuration by
\begin{equation}
   p =  (N_{\rm TP}, N_{\rm PP}),
\end{equation}
where $N_{\rm TP}$ denotes the tensor-parallel degree and $N_{\rm PP}$ the pipeline-parallel degree. Then, the per request total latency is given by
\begin{equation}\label{T_req}
    T_{\rm req}^{(p)} = T_{\rm prefill}^{(p)} + T_{\rm decode}^{(p)},
\end{equation}
where $T_{\rm prefill}^{(p)}$ and $T_{\rm decode}^{(p)}$
represent the prefill and decoding latency, respectively. Such decomposition immediately exposes a fundamental trade-off in LLM serving, where TTFT is primarily determined by the efficiency of prefill, whereas sustained token generation is determined by the efficiency of decoding. Consequently, optimizing a serving system for low TTFT does not necessarily produce the lowest per-token latency or highest token generation throughput. TP distributes the computation of each layer across $N_{\rm TP}$ GPUs, reducing the computation assigned to each GPU, while introducing collective communication. PP instead distributes model layers across $N_{\rm PP}$ consecutive stages, thereby reducing the amount of fine-grained inter-GPU synchronization, however, overheads related to pipeline scheduling and bubbles \footnote{Pipeline bubble refers to idle periods arising from inefficiencies when using PP to keep all stages continuously busy. Its impact depends on the amount of available workload and the scheduling policy to amortize this idle time \cite{mirzaei2026characterizingparallelismstrategiesllm}.} are inevitable \cite{mirzaei2026characterizingparallelismstrategiesllm}. We can therefore express the latency of a given parallelism configuration as follows
\begin{equation}\label{phase_comp}
    T_q^{(p)} = \tau_q^{(p)} + \eta_q^{(p)} + \gamma_q^{(p)},
\end{equation}
where $\tau_q^{(p)}$, $\eta_q^{(p)}$, and $\gamma_q^{(p)}$ represent computation, communication/synchronization, and scheduling overhead, respectively, assuming  parallelism configuration $p$, and  $q \in \{{\rm prefill, decode}\}$. Different parallelism configurations exhibit different interplay between the components in \eqref{phase_comp}. For example, with TP, increasing $N_{\rm TP}$ reduces the computation per GPU, while increasing the amount of inter-GPU communication. Thus, for a given phase $q$, the reduction in computation (i.e., $\tau_q^{(p)}$) may outweigh the communication cost (i.e., $\eta_q^{(p)}$). As the computation per iteration becomes smaller, however, communication becomes increasingly visible in the critical path. This effect is particularly important during decoding, where each iteration produces only a small number of new tokens. TP can therefore provide strong decode efficiency while potentially paying a larger communication cost during prefill. PP exhibits a complementary behavior. By assigning contiguous layers to different pipeline stages, PP avoids much of the fine-grained collective communication of TP, which makes it attractive for compute-intensive execution of prefill. However, its decode efficiency can be limited by pipeline scheduling and bubble effects, particularly when all stages are not fully occupied \cite{mirzaei2026characterizingparallelismstrategiesllm}. To characterize such competing effects, we represent the workload by
\begin{equation}
   w =  (L_{\rm in}, L_{\rm out}, C),
\end{equation}
where \(L_{\mathrm{in}}\) is the input sequence length (ISL), \(L_{\mathrm{out}}\) is the output sequence length (OSL), and \(C\) is the request concurrency. We also define $T_{\rm TTFT}^{(p,w)}$ and $T_{\rm ITL}^{(p,w)}$ random variables representing TTFT and ITL of a request served under parallelism $p$ and workload $w$, respectively. For given serving policy \(s\), model $\mathcal{M}$, hardware platform $\mathcal{H}$, parallelism configuration $p$, executing workload \(w\) produces a latency tuple\footnote{Here, we omit the dependency to $s$, $\mathcal{M}$ and $\mathcal{H}$ for ease of notations.}
\begin{equation} \label{ell}
\ell(p,w)=\left({\rm TTFT}^{\rm avg} (p,w), {\rm ITL}^{\rm avg} (p,w) \right),
\end{equation}
where ${\rm TTFT}^{\rm avg} (p,w) = {\rm mean}(T_{\rm TTFT}^{(p,w)})$\footnote{{\rm mean}$(\cdot)$ represents the mean function.} captures the average TTFT across all serving requests, whereas ${\rm ITL}^{\rm avg} (p,w) = {\rm mean}(T_{\rm ITL}^{(p,w)})$ captures the average sustained latency of autoregressive token generation. The tuple in \eqref{ell} provides two complementary views of inference latency. We deliberately retain these quantities as separate dimensions rather than combining them into a single metric, since their relative importance changes with the composition of the workload. For a request producing $L_{\mathrm{out}}$ tokens, the total request latency is given by \eqref{T_req}. Such decomposition highlights why a single latency measure can obscure important differences between parallelism configurations. In particular, for specific workload $w$, parallelism configuration $p_i$ and $p_j$ may have substantially different TTFT and ITL characteristics while producing similar total latency.
\subsection{Latency Feasibility Set and Pareto Frontier}
In order to characterize the latency feasibility set, we consider a constant total workload with fixed concurrency and total sequence length, denoted as $L_{\rm seq}$. We define 
\begin{equation}
\mathcal{W} = \{ (L_{\rm in}, L_{\rm out}, C) | L_{\rm in}+ L_{\rm out} = L_{\rm seq}\},
\end{equation}
as the set of feasible workloads, with 
\begin{equation}
\mathcal{P} = \{(N_{\rm TP}, N_{\rm PP}) | N_{\rm TP} N_{\rm PP} = N\},
\end{equation}
representing the set of available parallelism configurations, where $N$
denotes the total number of available GPUs. Note that fixing \(L_{\mathrm{seq}}\) defines a particular workload scale for the analysis. We fix the total sequence length to define a consistent workload scale, so that comparisons across different \(L_{\mathrm{in}}\) and \(L_{\mathrm{out}}\) values reflect changes in workload composition rather than changes in the overall request size. This allows us to construct the Pareto frontier for a well-defined workload family and isolate how the prefill and decode imbalance affects the latency trade-off. As we scale \(L_{\mathrm{seq}}\), the underlying workload changes and a new latency feasible set and Pareto frontier must consequently be constructed. 
It is also worth mentioning that, changing input and output lengths independently changes multiple aspects of the workload simultaneously. For example, increasing \(L_{\mathrm{in}}\), will increase the amount of prefill computation, while increasing \(L_{\mathrm{out}}\) will lead to more autoregressive decode iterations. If the total sequence length is also allowed to vary, differences in latency could instead be attributed simply to differences in the overall workload. By fixing \(L_{\mathrm{seq}}\), the approximate total workload stays constant while changing its composition between prefill and decode.
The corresponding workload-conditioned feasible set is therefore given by
\begin{equation}\label{region}
\mathcal{L}(w, p) = \{\ell(p, w) |w \in \mathcal{W}, p \in \mathcal{P}\}.
\end{equation}
For each specific workload \(w\), a Pareto frontier can be constructed independently. A configuration \(p_i\) dominates \(p_j\) under the workload \(w\) if
\begin{equation} 
{\rm TTFT}^{\rm avg}(p_i,w) \leq {\rm TTFT}^{\rm avg}(p_j,w) \quad \text{and} \quad {\rm ITL}^{\rm avg}(p_i,w) \leq {\rm ITL}^{\rm avg}(p_j,w), \end{equation}
with at least one inequality being strict. The Pareto-optimal configuration set is therefore
\begin{equation}
\mathcal{P}^*(w) = \{p \in \mathcal{P}  | \nexists p^{'} {\rm \, s.\, t.\,\,}  p^{'}\prec_w p\}.
\end{equation}
The corresponding TTFT–ITL Pareto frontier is given by
\begin{equation}\label{pareto}
\mathcal{F}(w) = \{\ell(p, w) \in \mathcal{L}(w, p)  | p \in \mathcal{P}^*(w)\}.
\end{equation}
Note that $\mathcal{F}(w)$ contains only those points for which no other configuration simultaneously provides lower TTFT and lower ITL. The frontier therefore represents the best achievable trade-offs between TTFT and ITL for that particular workload. The formulations in \eqref{region} and \eqref{pareto} provide a precise foundation for the notion of parallelism operating regions, i.e., the region of the workload space where different parallelism configurations occupy favorable positions on the latency frontier. This formulation further distinguishes two questions: (1) what TTFT–ITL combinations can be achieved, and (2) which configurations provide the best trade-offs between TTFT and ITL, such that improving one metric would require sacrificing the other. The first is captured by the feasible set \(\mathcal{L}(w)\), while the second is captured by the Pareto frontier \(\mathcal F(w)\).

For a given  $L_{\rm seq}$ and concurrency, we vary the input/output composition, construct the feasible latency set for every resulting workload, and extract its Pareto frontier. Note that feasible set and its corresponding Pareto frontier are workload-dependent. Therefore, the objective is not to identify a single globally optimal parallelism configuration, but to characterize how the TTFT–ITL trade-off and its Pareto-optimal configurations evolve with the workload. 
\subsection{Experiment Results on Pareto Frontier}
Here, we characterize the latency trade-off exposed by different parallelism configurations. For the purpose of evaluation, we employ Qwen2.5-14B-Instruct \cite{qwen2025qwen25technicalreport} with BF16 tensor type on $8\times$ NVDIA H200 ($141$GB) \cite{NvdaH200}. All experiments are performed using vLLM v0.15.0 as an inference engine \cite{vllm15}.  For the sake of descriptive notations throughout the experiments, we use the mapping in Table \ref{tab:mapping} for parallelism configuration across 8 GPUs
\begin{table}[h!]
    \centering
        \caption{Notation mapping.}
    \begin{tabular}{c|c|c|c|c}
    \hline
      $p$   &  $(8,1)$ &  $(4,2)$ & $(2,4)$  & $(1,8)$ \\
      \hline
       Notation & TP8  & TP4PP2  & TP2PP4 & PP8\\
        \hline
    \end{tabular}
    \label{tab:mapping}
\end{table}
% \begin{table}[h!]
%     \centering
%         \caption{Notation mapping.}
%     \begin{tabular}{c||c}
%     \hline
%       $p$   &  notation\\
%       \hline
%        $(8,1)$  & TP8\\
%        $(4,2)$  & TP4PP2\\
%        $(2,4)$  & TP2PP4\\
%        $(1,8)$  & PP8\\
%         \hline
%     \end{tabular}
%     \label{tab:mapping}
% \end{table}

\begin{figure}[t!]
    \centering
    \includegraphics[width=0.7\linewidth]{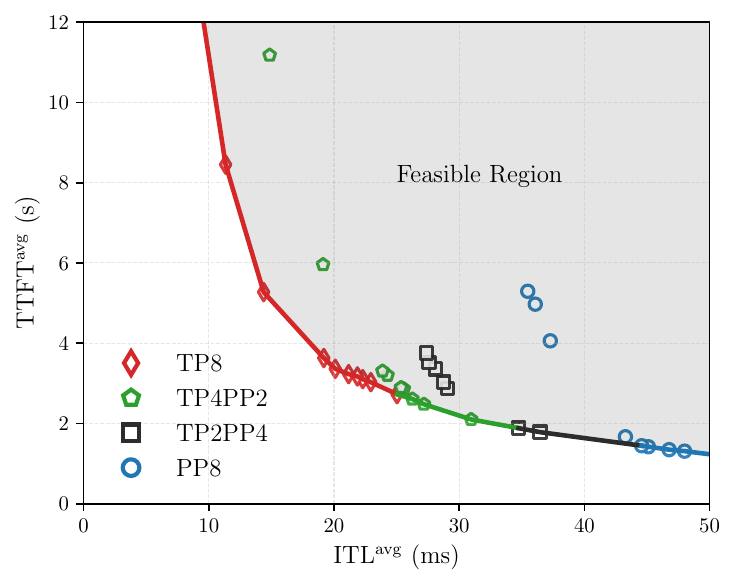}
    \caption{TTFT-ITL feasible region and Pareto frontier for different parallelism configurations. TP-dominant configurations achieve lower ITL but higher TTFT, while PP-dominant configurations achieve lower TTFT at the expense of higher ITL. As workloads transition from decode-heavy to prefill-heavy, the Pareto-optimal operating point shifts systematically from TP-dominant configurations, through hybrid configurations, toward PP-dominant configurations.}
    \label{fig:TTFT_vs_ITL_1}
\end{figure}
Fig. \ref{fig:TTFT_vs_ITL_1} depicts the latency feasible region and the corresponding Pareto frontier comparing different parallelism configurations. During inference, we consider batch size to be $128$, maximum GPU memory utilization of $0.8$ and prefill chunking is enabled. The rest of serving parameters remain as default values. We also consider a set of workloads $w=(L_{\rm in}, L_{\rm out}, 128)$, where $L_{\rm in}+ L_{\rm out} = 1200$. Indeed, the total sequence length is fixed, i.e.,  \(L_{\mathrm{seq}} = 1200\), while varying only the composition of the request by changing \(L_{\mathrm{in}}\) and \(L_{\mathrm{out}}\). Each point in the feasible region therefore represents the latency obtained by one parallelism configuration for a particular workload composition. For each parallelism configuration $p$, moving across the points corresponds to moving from decode-heavy workloads, with relatively small \(L_{\mathrm{in}}\) and large \(L_{\mathrm{out}}\), toward prefill-heavy workloads, with relatively large \(L_{\mathrm{in}}\) and small \(L_{\mathrm{out}}\). Since the total sequence length remains fixed, the figure isolates the effect of changing the relative contribution of prefill and decode rather than simply increasing the overall workload size. Therefore, the collection of points corresponding to the available parallelism configurations defines the latency feasible set corresponding to given workload. The resulting frontier reveals a clear and systematic transition across parallelism configurations. 

In Fig. \ref{fig:TTFT_vs_ITL_1}, we observe that TP8 occupies the low-ITL/high-TTFT end of the frontier, PP8 occupies the low-TTFT/high-ITL end, while the hybrid configurations provide intermediate trade-offs. The contrastive behavior of TP8 and PP8 can be explained by the fundamentally different compute-communication trade-off induced by TP and PP. During prefill, the model processes a relatively large amount of input computation in a highly parallel manner. TP distributes each layer's computation across GPUs, but requires collective communication between GPUs during the execution of the layers. Because prefill involves substantial activation processing, these communication operations can contribute significantly to the critical path. PP, in contrast, partitions the model into sequential stages and largely avoids the fine-grained collective communication required by TP. At sufficiently high concurrency, the pipeline can also remain well utilized. Consequently, PP8 can achieve substantially lower TTFT than TP8, placing PP8 toward the bottom of the figure.

The behavior is reversed during decode. Each decoding iteration processes only the newly generated token, making the computation per iteration relatively small. Under these conditions, the communication and synchronization costs associated with pipeline execution become more prominent relative to useful computation \cite{mirzaei2026characterizingparallelismstrategiesllm}. A deep PP configuration must propagate each token through multiple pipeline stages, and the limited amount of computation available per decoding step makes pipeline scheduling overhead more difficult to amortize. TP8, on the other hand, distributes the computation of every layer across all available GPUs, thereby reducing the computation required by each GPU and providing a highly parallel execution path for each decoding iteration. Although TP still incurs collective communication, its lower per-token computation time results in substantially lower ITL. TP8 therefore occupies the low-ITL end of the frontier.

This creates a TP-PP latency trade-off depicted in Fig. \ref{fig:TTFT_vs_ITL_1}. TP8 provides the fastest sustained token generation but incurs a higher cost to produce the first token. On the other hand, PP8 provides the fastest first response but sacrifices per-token generation latency. The hybrid configurations, TP2PP4 and TP4PP2, lie between these two extremes because they inherit the characteristics of both TP and PP. Increasing the TP degree progressively favors decode efficiency, while increasing the PP degree progressively favors prefill efficiency.

It is worth mentioning that, as workload transitions from decode-heavy to prefill-heavy, the Pareto-optimal operating point shifts systematically from TP-dominant configurations, through hybrid configurations, toward PP-dominant configurations. To explain this, consider a decode-heavy workload, characterized by a relatively small \(L_{\mathrm{in}}\) and large \(L_{\mathrm{out}}\). In this regime, the request spends comparatively more of its execution in autoregressive decoding. Consequently, the sustained per-token cost becomes increasingly important, and configurations with lower ITL provide more favorable latency behavior. We observe that TP8 configuration lies at the most favorable end of the frontier for this regime, making high-degree TP an effective configuration for decode-heavy workloads.

As \(L_{\mathrm{in}}\) increases and \(L_{\mathrm{out}}\) decreases, the workload becomes increasingly balanced between prefill and decode. Neither phase overwhelmingly determines the latency behavior, and the advantage of an extreme TP or PP configuration diminishes. The hybrid configurations consequently become increasingly attractive, occupying the middle portion of the Pareto frontier. Moreover, the two hybrid configurations provide a gradual transition between the TP- and PP-dominant regimes, i.e., TP4PP2 retains more of the TP characteristics and therefore favors lower ITL, whereas TP2PP4 shifts further toward the PP characteristics and favors lower TTFT.
\begin{figure}[t!]
    \centering
    \includegraphics[width=0.7\linewidth]{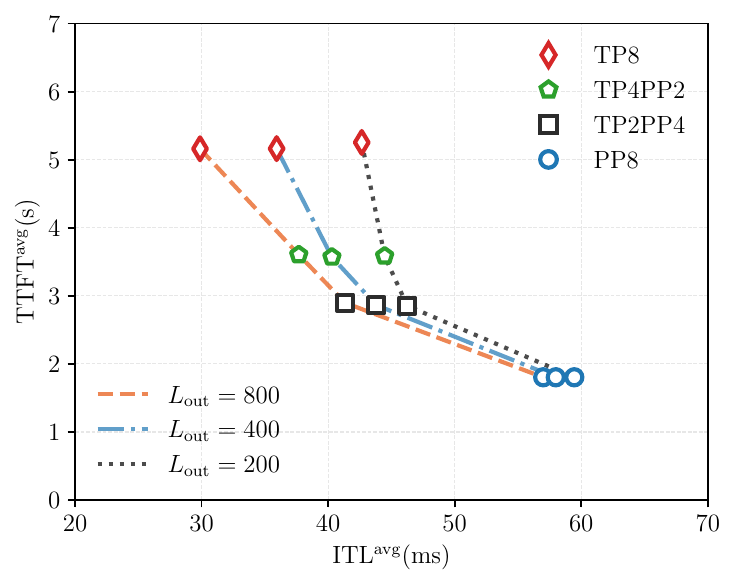}
    \caption{TTFT vs. ITL for various OSL. The effect of OSL is not uniform across parallelism configurations. Longer OSL mainly improves ITL through better amortization of decode overheads, with the largest benefit observed for TP-dominant configurations.}
    \label{fig:TTFT_vs_ITL_2}
\end{figure}
Finally, as \(L_{\mathrm{in}}\) becomes large relative to \(L_{\mathrm{out}}\), workload becomes prefill-heavy. The cost of processing the input context dominates the first token latency, and the advantage of PP becomes increasingly pronounced. PP8 therefore moves toward the favorable end of the latency trade-off, providing the lowest TTFT among the evaluated configurations, albeit with the highest ITL. In this regime, the additional decode efficiency of TP8 provides diminishing benefit because relatively little of the request is spent generating output tokens.

 Note that the four configurations should not be viewed as having a fixed global performance ranking. Instead, each configuration occupies a different position in the TTFT-ITL design space, and the desirable operating point changes as the workload moves through different prefill - decode regimes. The Pareto frontier provides a compact representation of this behavior by exposing the set of non-dominated latency trade-offs available for each workload.

In Fig. \ref{fig:TTFT_vs_ITL_2}, we consider the workload $w=(1000, L_{\mathrm{out}}, 256)$, and isolate the effect of OSL on the TTFT-ITL trade-off by varying \(L_{\mathrm{out}}\in\{200,400,800\}\). As \(L_{\mathrm{out}}\) increases, the latency frontier expands primarily along the ITL dimension, while its TTFT range remains largely unchanged. This behavior is expected because TTFT is dominated by the prefill phase, whose workload is determined primarily by the fixed input length and is therefore relatively insensitive to the number of tokens subsequently generated. In contrast, a longer $L_{\mathrm{out}}$ provides more decoding iterations over which per-request startup, scheduling, and pipeline effects can be amortized. Consequently, the effective measured ITL decreases as \(L_{\mathrm{out}}\) increases, exposing a wider range of achievable TTFT–ITL trade-offs. Thus, increasing OSL does not substantially change the TTFT, but allows the system to better amortize decode-related overheads over a longer generation process.

Importantly, the effect of OSL is not uniform across parallelism configurations. PP8 shows relatively little improvement in ITL as OSL increases, whereas the benefit becomes progressively larger for TP2PP4 and, most notably, TP8. This reflects the different strengths of the parallelism strategies during decoding. PP-dominant configurations incur pipeline-related synchronization and stage-propagation costs on each decoding iteration, mainly because these costs remain significant throughout generation, increasing the number of output tokens provides limited improvement in the measured per-token latency \cite{mirzaei2026characterizingparallelismstrategiesllm}. In contrast, TP-dominant configurations are better suited to autoregressive decoding. Once the decoding pipeline is established, the computation can be distributed across more GPUs and the associated per-request overheads are amortized over increasingly many decoding iterations. As a result, longer OSL increasingly favors TP-dominant configurations, causing their points to move toward lower ITL while their TTFT remains largely unchanged. The figure therefore shows that OSL does not merely shift the overall latency frontier; it changes the relative benefit of parallelism configurations, increasingly favoring TP as the workload becomes more decode-intensive.
\section{SLO-Aware Parallelism} \label{sec:SLO}
In the previous section, we characterized the fundamental latency trade-offs induced by different parallelism configurations through the TTFT-ITL Pareto frontier. This perspective is important because it reveals the inherent operating regions of TP-, PP-, and hybrid-parallel execution and identifies the set of non-dominated latency trade-offs available for a given workload. However, practical serving systems are rarely optimized for latency alone. Instead, they must deliver as much useful service capacity as possible while satisfying application-level latency requirements. A configuration with higher throughput is not necessarily preferable if it produces excessive latency for a significant fraction of requests, while a configuration with excellent latency may leave substantial system capacity underutilized. 

Moreover, latency requirements are inherently asymmetric. For example, reducing TTFT from $2$(s) to $1$(s) can be highly valuable for an interactive application, whereas the same improvement may provide little additional benefit once the application's responsiveness target has already been met. This motivates extending the latency trade-off analysis to an SLO-aware serving perspective, where parallelism configurations are evaluated according to how much useful throughput they can sustain under a given latency requirement. For a parallelism configuration \(p\), let \(\lambda_p (w)\) denote its achieved request throughput for workload $w$. Our goal is to find the optimal parallelism configuration that maximizes the throughput while respecting the TTFT constraints \footnote{There are various criteria to choose the SLO constraints, such as decoding latency or end-to-end latency. Here, we consider TTFT as a SLO constraint.}. To do so, let \(F_p(t|w)\) denote the empirical CDF of $T_{\rm TTFT}^{(p,w)}$, i.e., \(F_p(t|w)=\Pr(T_{\rm TTFT}^{(p,w)}\leq t)\). In particular, $F_p(t|w)$ represents the fraction of the requests that satisfies the latency requirement for parallelism configuration $p$ and workload $w$. Assuming $T_s$ to be the target TTFT SLO constraint, goodput is defined as
\begin{equation} \label{goodput_eq}
G_p(w,T_s) = \lambda_p(w)F_p(T_s| w).
\end{equation}
Thus, goodput can be interpreted as the product of two competing properties of a parallelism configuration: I) the raw serving capacity \(\lambda_p (w)\) and II) the probability of satisfying the latency requirement \(F_p(T_s|w)\). Such decomposition is important because maximizing raw throughput alone can favor a configuration that could generate many requests but violates the TTFT requirement for a substantial fraction of them. As we later show in this section, and under tight SLO requirement, each parallelism exhibits a trade-off between these components. The goal is to choose a parallelism configuration that delivers the largest amount of useful serving capacity for a given latency requirement. Therefore, we formalize SLO-aware parallelism selection as follows
\begin{equation}\label{optimal_p_SLO}
p^*(w, T_s) =  \argmax_{\ p \in \mathcal{P}} G_p (w, T_s).
\end{equation}
 According to \eqref{optimal_p_SLO}, each configuration \(p\in\mathcal{P}\) contributes to goodput according to both its achievable request throughput \(\lambda_p(w)\) and the fraction of requests that satisfy the SLO, \(F_p( T_s|w)\). The optimization therefore does not simply favor the configuration with the highest throughput or the lowest TTFT; instead, it balances the two through their product. As \(T_s\) varies, the relative contribution of SLO compliance also changes, causing the maximizing configuration \(p^*(w, T_s)\) to potentially change. Therefore, the optimization in \eqref{optimal_p_SLO} tries to capture the best parallelism configuration that makes the most effective use of the available hardware for a given level of latency requirement.

Note that, when solving  \eqref{optimal_p_SLO}, there is a fundamental trade-off between latency compliance and serving capacity across different parallelism configurations. To show this, for two configurations \(p_i\) and \(p_j\), the optimal \(T_s^*\) should solve $G_{p_i}(w, T_s^*) = G_{p_j}(w, T_s^*)$, or equivalently $\lambda_{p_i}(w)F_{p_i}(T_s^*|w) =\lambda_{p_j}(w)F_{p_j}(T_s^*|w)$, and therefore
\begin{equation}
\frac{F_{p_i}(T_s^*|w)}{F_{p_j}(T_s^*|w)} = \frac{\lambda_{p_j}(w)}{\lambda_{p_i}(w)}.
\end{equation}
For tight TTFT SLO, configurations with favorable TTFT distributions, such as PP-dominant configurations, can achieve substantially higher SLO attainment, which allows them to overcome their lower raw throughput. As the SLO is relaxed ($T_s \rightarrow \infty$), the compliance probability $F_p(T_s|w)\rightarrow 1 \, \forall p \in \mathcal{P}$, and therefore goodput  approaches to raw throughput, i.e., \(G_p(w, T_s)\rightarrow\lambda_p(w)\). Thus, parallelism configurations with higher serving capacity, such as TP-dominant configurations, eventually become preferable.

\subsection{Experiment Results on SLO-Aware Parallelism Selection}
\begin{figure}[t!]
    \centering
    
   \includegraphics[width=\linewidth]{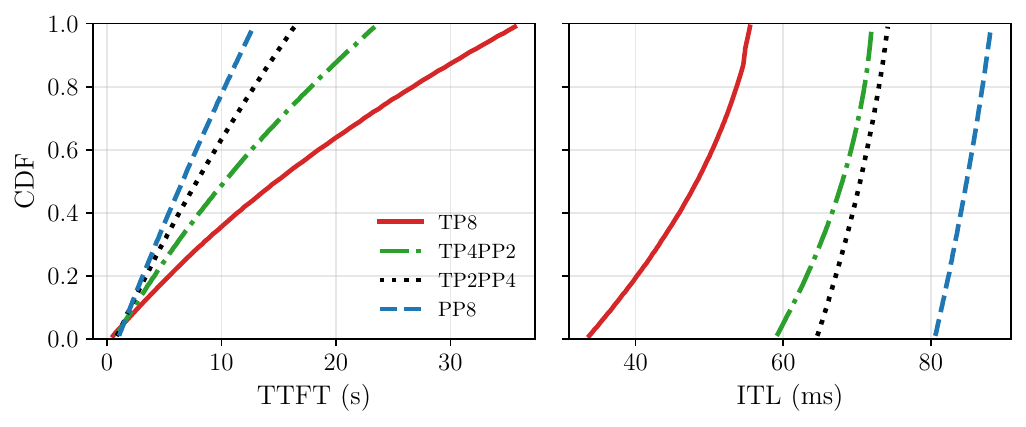}
        \caption{Left Plot: TTFT distribution (Prefill), Right plot: ITL distribution (decode). PP-dominant configurations achieve lower TTFT, while TP-dominant configurations achieve lower ITL, thereby highlighting the latency trade-off between prefill and decoding across different parallelisms.}
    \label{fig:TTFT_and_TPOT_CDF}
\end{figure}
Fig. ~\ref{fig:TTFT_and_TPOT_CDF} demonstrates the latency trade-offs by examining the distribution of request latency. In this experiment, we fix the workload as $w=(1200 ,2800, 256)$ where $L_{\rm seq} = 4000$. The left plot in Fig.~\ref{fig:TTFT_and_TPOT_CDF} shows the TTFT CDF during prefill, where PP8 provides the most favorable distribution, with its CDF rising more rapidly than the other configurations. This indicates that a larger fraction of requests achieve a lower TTFT under PP8. This is mainly due to that fact that prefill is compute-intensive, and PP partitions the model across stages while avoiding the fine-grained collective communication required by TP. As a result, PP8 can perform efficiently during the prefill, particularly when the amount of computation is sufficient to amortize pipeline overhead. Conversely, as we move towards TP-dominant configurations, such as TP8, they tend to exhibit a wider and slower-rising TTFT distribution, reflecting the additional communication associated with distributing each layer's computation across all GPUs. It is worth noting that hybrid configurations  provide intermediate behavior, which reinforces the observation that increasing the PP degree tends to favor TTFT.

The right plot in Fig. ~\ref{fig:TTFT_and_TPOT_CDF} shows the opposite behavior during autoregressive decoding. Here, TP8 provides the most favorable ITL distribution, with its CDF shifted substantially toward lower values, while PP8 exhibits wider ITL distribution. During decoding, each iteration processes only a small amount of new computation, making communication and synchronization overhead more significant relative to useful computation. TP8 can exploit all available GPUs to reduce the per-token computation time, thereby making it particularly effective for sustained token generation, whereas the sequential propagation of tokens through multiple pipeline stages makes PP8 less effective in this regime. The hybrid configurations lie between these two extremes, with their behavior determined by the balance between TP and PP. Together, the two CDFs provide a phase-level explanation for the TTFT–ITL Pareto frontier shown in Fig. \ref{fig:TTFT_vs_ITL_1}, where PP-dominant configurations favor the prefill phase and therefore TTFT, whereas TP-dominant configurations favor the decode phase and therefore ITL. 

\begin{figure}[t!]
    \centering
    \includegraphics[width=0.6\linewidth]{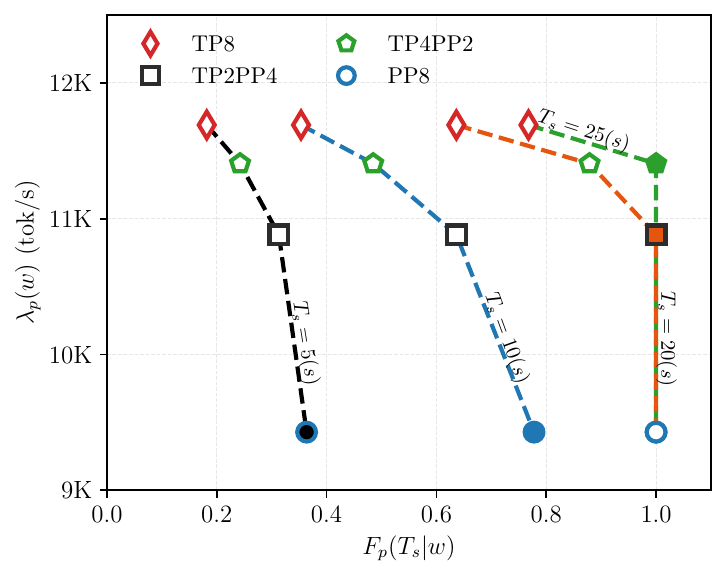}
    \caption{ Latency compliance and serving capacity trade-off for fixed workload $w=(1200 ,2800, 256)$ with $L_{\rm seq} = 4000$. Filled marker in each $T_s$, indicates the parallelism configuration that yields the maximum goodput for the corresponding $T_s$.}
    \label{fig:lambda_vs_CDF}
\end{figure}

Fig. \ref{fig:lambda_vs_CDF} provides a more direct decomposition of the SLO-aware goodput trade-off by plotting each configuration in the throughput-SLO-attainment space. Recall that goodput is \(G_p(T_s)=\lambda_pF_p(T_s)\)\footnote{Here, we assume workload $w$ is fixed and we omit its dependency.}, so a configuration must simultaneously provide high serving throughput \(\lambda_p\) and high SLO attainment \(F_p(T_s)\). As shown in Fig. \ref{fig:lambda_vs_CDF}, these two quantities are fundamentally in tension. Configurations with higher throughput (i.e., TP-dominant) generally achieve a smaller fraction of requests within a given TTFT SLO, while configurations with stronger TTFT performance (PP-dominant) tend to sacrifice throughput for a better TTFT SLO attainment. This occurs because the parallelism choices that accelerate prefill and reduce TTFT, particularly PP-dominant configurations, do so at the expense of decode efficiency and therefore sustained request throughput; TP-dominant configurations exhibit the opposite behavior \cite{mirzaei2026characterizingparallelismstrategiesllm}. Consequently, maximizing raw throughput alone is insufficient for maximizing goodput. 

More importantly, this observation makes the SLO-dependent shift in the optimal parallelism explicit. For each \(T_s\), the filled marker identifies the configuration maximizing the goodput, illustrating that the optimal point is determined by the balance between throughput capacity and SLO compliance rather than either quantity independently. Under stringent TTFT requirements (\(T_s=5\) and \(10\)(s)), SLO attainment is the dominant differentiating factor. PP8 achieves substantially higher \(F_p(T_s)\), which compensates for its lower throughput and results in the highest goodput. As the SLO is progressively relaxed to \(20\) and \(25\)(s), the SLO-attainment advantage of PP8 diminishes as more requests from all configurations satisfy the requirement. The throughput advantage of TP-dominant configurations can then dominate, shifting the goodput optimality toward the hybrid configurations and ultimately toward TP8. Thus, the plot reveals a fundamental capacity–compliance trade-off induced by parallelism. In particular, PP-dominant configurations trade throughput for stronger latency compliance, whereas TP-dominant configurations trade latency compliance for higher serving capacity. The optimal strategy consequently moves from PP to hybrid to TP as the SLO is relaxed, reinforcing the earlier observation for the operating-region in Fig. \ref{fig:TTFT_vs_ITL_1}.

\begin{figure}[t!]
    \centering
    \includegraphics[width=0.6\linewidth]{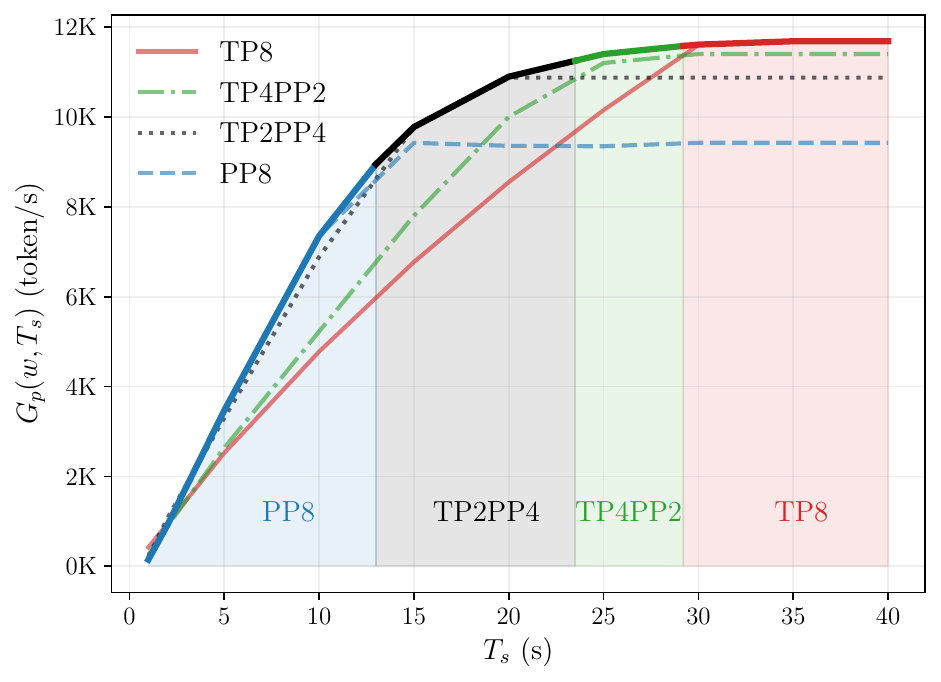}
    \caption{Goodput vs. TTFT SLO for fixed workload $w=(1200 ,2800, 256)$ where $L_{\rm seq} = 4000$. As TTFT SLO is relaxed, the optimal parallelism shifts from PP-dominant configurations to TP-dominant configurations.}
    \label{fig:goodput_TTFT_SLO}
\end{figure}

Fig. \ref{fig:goodput_TTFT_SLO} depicts goodput vs. TTFT SLO for various parallelism configurations. The observation in this figure indicates how goodput captures the interplay between a configuration's serving capacity and its ability to satisfy a latency requirement as shown in Fig. \ref{fig:lambda_vs_CDF}. Under a tight TTFT SLO, we observe that PP8 can achieve the highest goodput despite having a lower raw throughput \(\lambda_{\mathrm{PP8}}\) than TP-dominant configurations. The reason is that PP8 provides substantially higher SLO attainment, \(F_{\mathrm{PP8}}(T_s)\), because of its superior TTFT behavior. PP8 therefore wins whenever its compliance advantage is sufficiently large to compensate for its throughput disadvantage
\begin{equation}
   \lambda_{\rm PP8}F_{\rm PP8}(T_s) > \lambda_{\rm TP8}F_{\rm TP8}(T_s),
\end{equation}
or equivalently 
\begin{equation}
\frac{F_{\rm PP8}(T_s)}{F_{\rm TP8}(T_s)} > \frac{\lambda_{\rm TP8}}{\lambda_{\rm PP8}}.
\end{equation}
Thus, at stringent SLOs, the ability to meet the latency requirement matters more than raw serving capacity. For example, as shown in Fig. \ref{fig:lambda_vs_CDF}, TP8 may serve requests at a higher rate, but a large fraction of those requests may fail the TTFT requirement and consequently contribute less to goodput.
As the TTFT SLO is relaxed, such behavior progressively reverses. The SLO-attainment probability \(F_p(T_s)\) increases for all configurations and approaches one as the constraint becomes sufficiently loose. Consequently,
\begin{equation}
 \lim_{T_s \rightarrow \infty}G_p(T_s) = \lim_{T_s \rightarrow \infty} \lambda_p F_p(T_s) = \lambda_p,
\end{equation}
so goodput increasingly reflects the underlying throughput capacity of each configuration. The latency-compliance advantage of PP8 therefore becomes less valuable, while the higher decoding throughput of TP-dominant configurations becomes increasingly important. This produces the observed transition from PP8 at stringent SLOs, through hybrid configurations at intermediate SLOs, to TP8 under relaxed SLOs. Together with the latency trade-offs established earlier in Section \ref{sec:latency_pareto}, this result shows that the choice of parallelism is fundamentally SLO-dependent. PP-dominant parallelism is advantageous when latency compliance is the primary constraint, whereas TP-dominant parallelism becomes preferable when the latency requirement can be relaxed enough to exploit its greater serving capacity.

\section{Parallelism for Long Context} \label{sec:Long_Context}
The preceding sections characterize parallelism primarily through the latency trade-offs, and show that the preferred configuration depends on the workload composition and latency requirement. However, as LLM workloads move toward the long-context workloads, the ability of a parallelism configuration to accommodate the workload becomes critical. Increasing the input sequence length not only increases prefill computation and TTFT, but also increases the memory required to store the model's intermediate states and KV cache, potentially changing both performance and feasible operating range. Thus, a configuration that is attractive for shorter requests may become inefficient or even infeasible as context length grows. To isolate this effect, we fix the $L_{\rm out}$, and just increase $L_{\rm in}$, thereby increasing the request's context size while keeping the decode workload fixed. This setup allows us to study the evolution of throughput, latency, SLO compliance, and ultimately the feasible operating range of different parallelism configurations. In this setting, parallelism is no longer only a mechanism for trading TTFT against ITL; it also determines how far the system can scale in context length before performance degrades or the workload becomes infeasible.

\begin{figure}[t!]
    \centering
     \includegraphics[width=\linewidth]{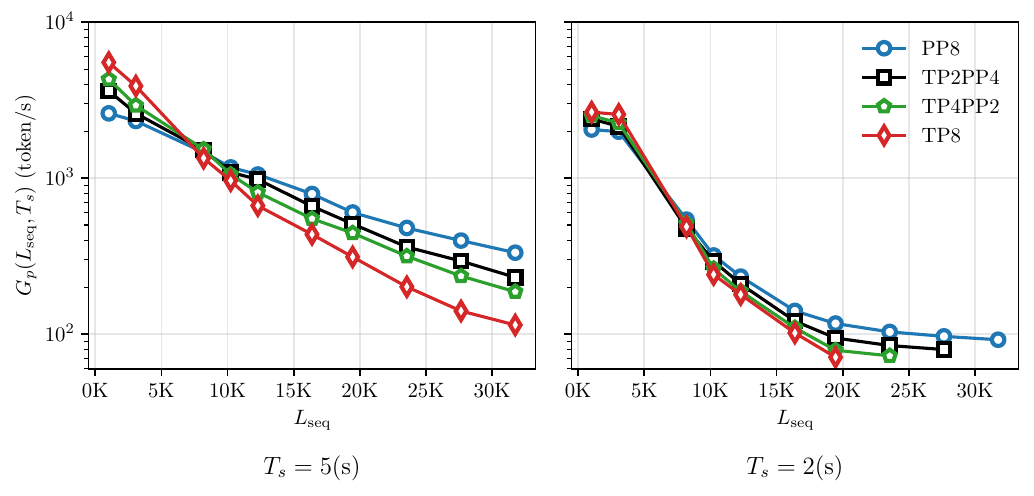}
    \caption{Goodput vs. Sequence Length for workload $w=(L_{\rm in} ,1000, 128)$ where $L_{\rm in}$ varies to construct values of x-axis. Left plot: $T_s = 5$(s). Right Plot: $T_s = 2$(s). Increasing sequence length progressively favors PP-dominant parallelism over TP-dominant parallelism, highlighting PP's stronger scalability for long context workloads.}
    \label{fig:goodput_Seqlen}
\end{figure}

Fig. \ref{fig:goodput_Seqlen} depicts the goodput against the sequence length for $T_s= 2$(s) and $5$(s). As sequence length increases, in both scenarios, we observe a systematic shift in the relative advantage of parallelism configurations toward PP-dominated strategies, indicating that PP-dominant parallelism scales more favorably with long-sequence workloads. In the left plot in Fig. \ref{fig:goodput_Seqlen}, where $T_s = 5$(s), at shorter sequence length, TP-dominant configurations outperform their PP-dominant counterparts, mainly due to their higher serving capacity and strong decode efficiency. However, as the input sequence length grows, prefill becomes increasingly dominant and the communication overhead associated with TP execution becomes more consequential. PP-dominant configurations, in contrast, avoid much of this fine-grained collective communication and therefore experience a slower degradation in TTFT and SLO attainment. Consequently, the initial throughput advantage of TP-dominant configurations is progressively eroded as sequence length increases, which allows PP-dominant configurations to become increasingly competitive and eventually achieve higher goodput.

\begin{figure}[t!]
    \centering
     \includegraphics[width=\linewidth]{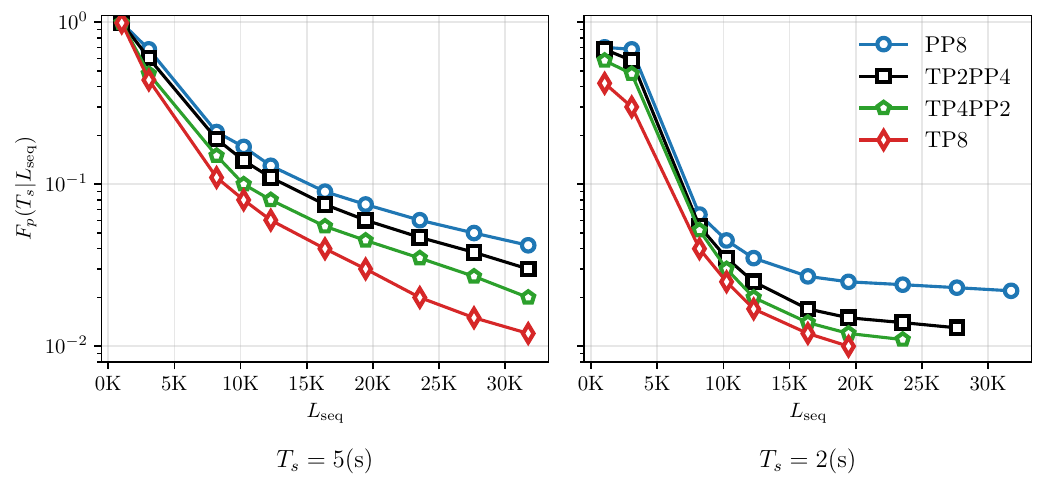}
    \caption{CDF vs. Sequence Length for workload $w=(L_{\rm in} , 1000, 128)$ where $L_{\rm in}$ varies to construct values of x-axis.  Left plot: $T_s = 5$(s). Right plot: $T_s = 2$(s). 
    PP-dominant configurations maintain higher SLO attainment as sequence length increases, especially under tight TTFT SLOs.}
    \label{fig:CDF_ctx_length}
\end{figure}

This effect is particularly pronounced under stringent TTFT requirements as indicated in the right plot in Fig. \ref{fig:goodput_Seqlen}, where $T_s = 2 (s)$. From \eqref{goodput_eq}, where \(G_p(w, T_s)=\lambda_p(w)F_p(T_s|w)\), we note that long-context scaling is determined not only by the throughput \(\lambda_p(w)\), but also by how rapidly SLO attainment \(F_p(T_s|w)\) deteriorates with increasing sequence length. Our results indicate that PP-dominant configurations maintain higher SLO attainment over a wider range of sequence lengths (illustrated in Fig. \ref{fig:CDF_ctx_length}), which enables them to sustain useful throughput farther into the long-context regime. Thus, the key observation is not simply that PP performs better at the largest sequence length, but that PP-dominant parallelism exhibits more favorable scaling with sequence length, with its advantage becoming increasingly important as the TTFT SLO becomes tighter.

% \begin{figure}[t!]
%     \centering
    
%     \begin{subfigure}[t]{0.49\textwidth}
%         \centering
%         \includegraphics[width=\linewidth]{figs/goodput_Seqlen_5s.pdf}
%         \caption{$T_s= 5 {\rm (s)}$}
%         \label{fig:goodput_Seqlen_5s}
%     \end{subfigure}
%     \hfill
%     \begin{subfigure}[t]{0.49\textwidth}
%         \centering
%         \includegraphics[width=\linewidth]{figs/goodput_Seqlen_2s.pdf}
%         \caption{$T_s= 2 {\rm (s)}$}
%         \label{fig:goodput_Seqlen_2s}
%     \end{subfigure}

%     \caption{Goodput vs Sequence Length - OSL = 1000, Concurrency = 128.}
%     \label{fig:goodput_Seqlen}
% \end{figure}

% \begin{figure}[t!]
%     \centering
    
%     \begin{subfigure}[t]{0.49\textwidth}
%         \centering
%         \includegraphics[width=\linewidth]{figs/CDF_ctx_length_5.pdf}
%         \caption{$T_s= 5 {\rm (s)}$}
%         \label{fig:CDF_ctx_length_5}
%     \end{subfigure}
%     \hfill
%     \begin{subfigure}[t]{0.49\textwidth}
%         \centering
%         \includegraphics[width=\linewidth]{figs/CDF_ctx_length_2.pdf}
%         \caption{$T_s= 2 {\rm (s)}$}
%         \label{fig:CDF_ctx_length_2}
%     \end{subfigure}
%     \caption{CDF vs Sequence Length - OSL = 1000, Concurrency = 128.}
%     \label{fig:CDF_ctx_length}
% \end{figure}
Fig.~\ref{fig:CDF_ctx_length} isolates the SLO-attainment component of goodput and shows the sensitivity of $F_p(T_s|w)$ with respect to sequence length under both tight and relaxed TTFT SLO requirement. In particular, for each parallelism configuration we plot\footnote{In this experiment, $L_{\rm out}$ and $C$ is fixed. Therefore, dependency to $w$ boils down to $L_{\rm seq}$ only.}
\begin{equation}
    F_p(T_s|L_{\rm seq}) =  Pr(T_{\rm TTFT}^{(p,L_{\rm seq})} < T_s | L_{\rm seq}).
\end{equation}
In the left plot of Fig. \ref{fig:CDF_ctx_length}, as the sequence length increases, \(F_p(T_s\mid L_{\rm seq})\) consistently decreases across all configurations. This reflects the growing cost of processing the input sequence during prefill. Longer sequence increases the work required before the first token can be produced, thereby shifting the TTFT distribution toward larger values and causing an increasing fraction of requests to violate the required TTFT SLO. This effect becomes substantially sharper for the tighter \(T_s=2\)(s) as shown in the right plot of Fig. \ref{fig:CDF_ctx_length}. Intuitively, a relaxed SLO lies farther into the right side of the TTFT distribution, so moderate increases in TTFT with sequence length may leave many requests still compliant. In contrast, a tighter threshold lies much deeper in the low-latency portion of the distribution; where even a modest shift in TTFT therefore causes a large reduction in the fraction of requests satisfying the TTFT SLO. Thus, tighter latency requirements amplify the impact of sequence length on SLO attainment.

The observation in Fig. ~\ref{fig:CDF_ctx_length} further indicates that different parallelism configurations exhibit different sequence length scaling behavior. The relevant quantity is not only the absolute TTFT at a particular sequence length, but how rapidly SLO attainment degrades as sequence length grows. In particular, the PP-dominant configurations retain a larger \(F_p(T_s\mid L_{\rm seq})\) over increasing sequence lengths, while TP8 exhibits a substantially faster decline, especially under the tighter SLO. This difference directly propagates to goodput through
\begin{equation}
    G_p( L_{\rm seq}, T_s) = \lambda_p(L_{\rm seq})F_p(T_s|L_{\rm seq}).
\end{equation}
Although TP-dominant configurations may have higher throughput \(\lambda_p\), that advantage can be progressively overwhelmed when their SLO-attainment probability falls more rapidly with sequence length. Conversely, PP-dominant configurations can sustain a larger fraction of SLO-compliant requests as the sequence length grows, thereby allowing their goodput to degrade more gracefully and eventually surpass TP-dominant configurations. The key insight is that long-sequence scalability is governed not only by throughput degradation, but also by the rate at which latency-SLO attainment deteriorates with sequence length. This provides direct explanation for the increasing advantage of PP-dominant parallelism observed in Fig. \ref{fig:goodput_Seqlen}, particularly under stringent TTFT requirements.

% \begin{figure}[t!]
%     \centering
%     \begin{subfigure}[t]{0.49\textwidth}
%         \centering
%         \includegraphics[width=\linewidth]{figs/TTFT_P99_Long.png}
%         \label{fig:TTFT_P99_Long}
%     \end{subfigure}
%     \hfill
%     \begin{subfigure}[t]{0.49\textwidth}
%         \centering
%         \includegraphics[width=\linewidth]{figs/TPOT_P99_Long.png}
%         \label{fig:TPOT_P99_Long}
%     \end{subfigure}

%     \caption{P99 Latency vs Sequence Length}
%     \label{fig:goodput_Seqlen.}
% \end{figure}

% \begin{figure}[h!]
%     \centering
%     \begin{subfigure}[t]{0.49\textwidth}
%         \centering
%         \includegraphics[width=\linewidth]{figs/Prefill_through_Long.png}
%         % \caption{Prefill Throughput}
%         \label{fig:Prefill_through_Long}
%     \end{subfigure}
%     \hfill
%     \begin{subfigure}[t]{0.49\textwidth}
%         \centering
%         \includegraphics[width=\linewidth]{figs/Decode_through_Long.png}
%         % \caption{Decoding Throughput}
%         \label{fig:Decode_through_Long}
%     \end{subfigure}

%     \caption{Prefill and decoding Throughput vs Sequence Length}
%     \label{fig:PD_Throughputt_Seqlen.}
% \end{figure}

\section{Conclusions} \label{sec:Conclusions}
In this paper, we presented a systematic study of LLM inference parallelism from the perspectives of latency trade-offs, SLO-aware serving, and long-context scalability. We showed that TP, PP, and their hybrid combinations occupy distinct operating points in the TTFT-ITL design space, resulting in a workload-dependent Pareto frontier rather than a single globally optimal configuration. Our analysis revealed a clear transition from TP-dominant configurations for decode-heavy workloads to PP-dominant configurations for prefill-heavy workloads, which highlights the importance of considering workload composition when selecting inference parallelism. We further demonstrated that average latency metrics alone are insufficient to characterize serving performance and that TTFT and ITL should be treated as complementary dimensions when evaluating distributed inference systems.

Building upon this latency-centric view, we formulated SLO-aware parallelism selection through goodput and showed that the optimal configuration depends jointly on serving capacity and latency compliance. Our results indicate that stringent TTFT requirements favor PP-dominant configurations, whereas relaxed SLOs progressively shift the optimality toward TP-dominant configurations with higher throughput capacity. Finally, we demonstrated that increasing context length fundamentally changes the relative advantages of parallelism strategies, with PP-dominant configurations exhibiting more favorable scaling and maintaining higher SLO attainment in the long-context regime. Collectively, these findings suggest that parallelism should be treated not as a fixed deployment choice, but as a workload-, SLO-, and context-dependent operating decision. We believe that the proposed operating-region framework provides a useful foundation for future adaptive serving systems capable of dynamically selecting parallelism configurations in response to changing workload conditions and application requirements.

\section*{Acknowledgments} 
The authors would like to acknowledge the support of the Dell leadership team and thank colleagues across Dell for their valuable discussions, constructive feedback, and technical insights that contributed to this research.

\bibliographystyle{ieeetr}  
\bibliography{references}  %%% Remove comment to use the external .bib file (using bibtex).

@inproceedings{agrawal2024taming,
  title={Taming Throughput-Latency tradeoff in {LLM} inference with Sarathi-Serve},
  author={Agrawal, A. and Kedia, N. and Panwar, A. and Mohan, J. and Kwatra, N. and Gulavani, B. and Tumanov, A. and Ramjee, R.},
  booktitle={18th USENIX symp. on operating sys. design and implementation (OSDI 24)},
  pages={117--134},
  year={2024}
}

@misc{wang2026tamingrequestimbalancesloaware,
      title={Taming Request Imbalance: {SLO}-Aware Scheduling for Disaggregated {LLM} Inference}, 
      author={Qipeng Wang},
      year={2026},
      note = {Available:{https://arxiv.org/abs/2605.02329}}, 
}

@inproceedings{zhong2024distserve,
  title={{DistServe}: Disaggregating prefill and decoding for goodput-optimized large language model serving},
  author={Zhong, Y. and Liu, S. and Chen, J. and Hu, J. and Zhu, Y. and Liu, X. and Jin, X. and Zhang, H.},
  booktitle={18th USENIX Symposium on Operating Systems Design and Implementation (OSDI 24)},
  pages={193--210},
  year={2024}
}

@misc{adnan2026cascadeexploitingsloawarelatency,
  title = {Cascade: Exploiting {SLO}-Aware latency budget for fair and high goodput {LLM} inference serving},
  author = {M. Adnan and R. Mahapatra and P. J. Nair and D. Berger and P. Zardoshti and R. Fonseca and E. Choukse},
  year = {2026},
  note = {Available:  {https://arxiv.org/abs/2608.06557}}
}

@misc{agrawal2024etalonholisticperformanceevaluation,
  title = {Etalon: Holistic Performance Evaluation Framework for {LLM} Inference Systems},
  author = {A. Agrawal and A. Agarwal and N. Kedia and J. Mohan and S. Kundu and N. Kwatra and R. Ramjee and A. Tumanov},
  year = {2024},
  note = {Available: {https://arxiv.org/abs/2407.07000}}
}

@misc{du2026goodservehighgoodputservingagentic,
  title = {{GoodServe}: Towards High-Goodput Serving of Agentic {LLM} Inferences over Heterogeneous Resources},
  author = {B. Du and B. Huangfu and Y. Luo and C. Chen and Z. Li and M. Yu and X. Fan and M. Guo},
  year = {2026},
  note = {Available:  {https://arxiv.org/abs/2605.16867}}
}

@misc{shi2025nexusproactiveintragpudisaggregationprefill,
  title = {Nexus: Proactive Intra-{GPU} Disaggregation of Prefill and Decode in {LLM} Serving},
  author = {X. Shi and C. Cai and J. Du and Z. Jia},
  year = {2025},
  note = {Available:  {https://arxiv.org/abs/2507.06608}}
}

@misc{wang2025prefilldecodeaggregationdisaggregationunifying,
  title = {Prefill-Decode Aggregation or Disaggregation? Unifying Both for Goodput-Optimized {LLM} Serving},
  author = {C. Wang and P. Zuo and Z. Chen and Y. Liang and Z. Yu and M. Yang},
  year = {2025},
  note = {Available: {https://arxiv.org/abs/2508.01989}}
}

@misc{goel2025niyamabreakingsilos,
  title = {{Niyama}: Breaking the Silos of {LLM} Inference Serving},
  author = {K. Goel and J. Mohan and N. Kwatra and R. S. Anupindi and R. Ramjee},
  year = {2025},
  note = {Available: {https://arxiv.org/abs/2503.22562}}
}

@misc{lin2024infinitellmefficientllmservice,
  title = {{Infinite-LLM}: Efficient {LLM} Service for Long Context with DistAttention and Distributed {KVCache}},
  author = {B. Lin and C. Zhang and T. Peng and H. Zhao and W. Xiao and M. Sun and A. Liu and Z. Zhang and L. Li and X. Qiu and S. Li and Z. Ji and T. Xie and Y. Li and W. Lin},
  year = {2024},
  note = {Available:  {https://arxiv.org/abs/2401.02669}}
}

@misc{zhang2025jengaeffectivememorymanagemen,
  title = {{Jenga}: Effective Memory Management for Serving {LLM} with Heterogeneity},
  author = {C. Zhang and K. Du and S. Liu and W. Kwon and X. Mo and Y. Wang and X. Liu and K. You and Z. Li and M. Long and J. Zhai and J. Gonzalez and I. Stoica},
  year = {2025},
  note = {Available:  {https://arxiv.org/abs/2503.18292}}
}

@misc{ruan2025dynaserveunifiedelasticexecution,
      title={{DynaServe}: Unified and Elastic Execution for Dynamic Disaggregated {LLM} Serving}, 
      author={C. Ruan and Y. Chen and D. Tian and Y. Shi and Y. Wu and J. Li and C. Li},
      year={2025},
      note = {Available: {https://arxiv.org/abs/2504.09285}}, 
}

@misc{zhou2025sparseserveunlockingparallelismdynamic,
  title = {{SparseServe}: Unlocking Parallelism for Dynamic Sparse Attention in Long-Context {LLM} Serving},
  author = {Q. Zhou and P. Yin and P. Zuo and J. Cheng},
  year = {2025},
  note = {Available: {https://arxiv.org/abs/2509.24626}}
}

@misc{shi2026vppvirtualpipelineparallelism,
  title = {{VPP}: Virtual Pipeline Parallelism for Efficient Chunked Prefill in Long-Context {LLM} Inference},
  author = {Y. Shi and X. Wang and J. Gao and J. Luo and X. Zhou and F. Liu and K. Luo and X. Li and X. Guo and L. Feng},
  year = {2026},
  note = {Available:  {https://arxiv.org/abs/2608.26523}}
}

@misc{zhang2025pecschedpreemptiveefficientcluster,
  title = {{PecSched}: Preemptive and Efficient Cluster Scheduling for {LLM} Inference},
  author = {Z. Zhang and H. Shen},
  year = {2025},
  note = {Available: {https://arxiv.org/abs/2409.15104}}
}

@misc{ghadia2026untiedulyssesmemoryefficientcontext,
  title = {{Untied Ulysses}: Memory-Efficient Context Parallelism via Headwise Chunking},
  author = {R. Ghadia and M. Abraham and S. Vorobyov and M. Ryabinin},
  year = {2026},
  note = {Available:  {https://arxiv.org/abs/2602.21196}}
}

@misc{agrawal2025medha,
  title = {No Request Left Behind: Tackling Heterogeneity in Long-Context {LLM} Inference with {Medha}},
  author = {Agrawal, A. and others},
  year = {2025},
  note = {Available:  {https://arxiv.org/abs/2409.17264}}
}

@misc{guo2026adaptivecontextparallelismproduction,
  title = {Adaptive Context Parallelism for Production {LLM} Serving},
  author = {J. Guo and R. Wang and P. Huang and Z. Lv and Z. Wang and K. Liu and T. Lan and L. Qu and X. Wang and T. Yang},
  year = {2026},
  note = {Available:  {https://arxiv.org/abs/2609.04774}}
}

@misc{song2026deadlineawareadaptiveprefillchunking,
      title = {Deadline-Aware Adaptive Prefill Chunking for Efficient Large Language Model Serving}, 
      year = {2026},
      note = {Available:  {https://arxiv.org/abs/2609.07883}}, 
}

@misc{cheng2026robustkvcachemanagement,
  title = {Robust {KV} Cache Management for {LLM} Serving under Output Token Length Uncertainty},
  author = {J. Cheng and D. T. Do and D. T. Nguyen},
  year = {2026},
  note = {Available:  {https://arxiv.org/abs/2607.16892}}
}

@misc{chen2025slosserveoptimizedservingmultislo,
  title = {{SLOs-Serve}: Optimized Serving of Multi-{SLO} {LLMs}},
  author = {S. Chen and Z. Jia and S. Khan and A. Krishnamurthy and P. B. Gibbons},
  year = {2025},
  note = {Available:  {https://arxiv.org/abs/2504.08784}}
}

@misc{srivatsa2026nitsumservingtieredllm,
  title = {Nitsum: Serving Tiered {LLM} Requests with Adaptive Tensor Parallelism},
  author = {V. Srivatsa and Z. He and P. Guo and D. Li and Y. Zhang},
  year = {2026},
  note = {Available:  {https://arxiv.org/abs/2605.05467}}
}

@misc{yuan2026remplowdowntimeruntimemodelparallelism,
  title = {{ReMP}: Low-Downtime Runtime Model-Parallelism Reconfiguration for {LLM} Serving},
  author = {H. Yuan and K. Zheng and Y. Bai and Y. Zhang and Y. Zhang and B. Wu and X. Gao and D. Cheng},
  year = {2026},
  note = {Available:  {https://arxiv.org/abs/2606.18741}}
}

@misc{gao2026flyingservingontheflyparallelis,
  title = {{FLYING SERVING}: On-the-Fly Parallelism Switching for Large Language Model Serving},
  author = {S. Gao and J. Yin and F. Wang and W. Dong},
  year = {2026},
  note = {Available:  {https://arxiv.org/abs/2602.22593}}
}

@misc{kwon2023efficientmemorymanagementlarge,
  title = {Efficient Memory Management for Large Language Model Serving with PagedAttention},
  author = {W. Kwon and Z. Li and S. Zhuang and Y. Sheng and L. Zheng and C. H. Yu and J. E. Gonzalez and H. Zhang and I. Stoica},
  year = {2023},
  note = {Available:  {https://arxiv.org/abs/2309.06180}}
}

@misc{amer2026distributedhybridparallelismlarge,
  title = {Distributed Hybrid Parallelism for Large Language Models: Comparative Study and System Design Guide},
  author = {H. Amer and R. Karim and A. Pourranjbar and W. Zhang and W. Ahmed and B. Chen},
  year = {2026},
  note = {Available:  {https://arxiv.org/abs/2602.09109}}
}

@misc{xu2025characterizingcommunicationpatternsdistributed,
  title = {Characterizing Communication Patterns in Distributed Large Language Model Inference},
  author = {L. Xu and K. K. Suresh and Q. Anthony and N. Alnaasan and D. K. Panda},
  year = {2025},
  note = {Available:  {https://arxiv.org/abs/2507.14392}}
}

@misc{hidayetoglu2026shiftparallelismlowlatencyhighthroughput,
      title = {Shift Parallelism: Low-Latency, High-Throughput {LLM} Inference for Dynamic Workloads}, 
      author = {M. Hidayetoglu and A. Qiao and M. Wyatt and J. Rasley and Y. He and S. Rajbhandari},
      year = {2026},
      note = {Available:  {https://arxiv.org/abs/2509.16495}}, 
}

@misc{guo2025gllmglobalbalancedpipeline,
  title = {{gLLM}: Global Balanced Pipeline Parallelism System for Distributed {LLM} Serving with Token Throttling},
  author = {T. Guo and X. Zhang and J. Du and Z. Chen and N. Xiao and Y. Lu},
  year = {2025},
  note = {Available:  {https://arxiv.org/abs/2504.14775}}
}

@misc{niu2026dhpefficientscalingmllm,
      title={{DHP}: Efficient Scaling of {MLLM} Training with Dynamic Hybrid Parallelism}, 
      author={Y. Niu and H. Xiao and D. Liu and W. Zhou and J. Li},
      year={2026},
      eprint={2602.21788},
      archivePrefix={arXiv},
      primaryClass={cs.DC},
      note = {Available: {https://arxiv.org/abs/2602.21788}}, 
}

@misc{zhao2026unleashingscalablecontextparallelism,
      title={Unleashing Scalable Context Parallelism for Foundation Models Pre-Training via {FCP}}, 
      author={Y. Zhao and X. Nie and K. Zhu and S. Ma and Z. Lai and H. Hao and Y. Zhou and B. Kasikci and I. Stoica},
      year={2026},
      eprint={2605.08524},
      archivePrefix={arXiv},
      primaryClass={cs.DC},
      note = {Available: {https://arxiv.org/abs/2605.08524}}, 
}

@misc{wang2026flashcploadbalancedcommunicationefficientcontext,
      title={{FlashCP}: Load-Balanced Communication-Efficient Context Parallelism for {LLM} Training}, 
      author={Z. Wang and E. Liu and L. Jiang and Z. Yu and Z. Pan and Y. Guan and Y. Wang and Y. Ding},
      year={2026},
      eprint={2606.08476},
      archivePrefix={arXiv},
      primaryClass={cs.DC},
      note = {Available: {https://arxiv.org/abs/2606.08476}}, 
}

@misc{jiang2025dcpaddressinginputdynamism,
      title={{DCP}: Addressing Input Dynamism In Long-Context Training via Dynamic Context Parallelism}, 
      author={C. Jiang and Z. Cai and Y. Tian and . Jia and Y. Wang and C. Wu},
      year={2025},
      eprint={2510.10620},
      archivePrefix={arXiv},
      primaryClass={cs.DC},
      doi={https://doi.org/10.1145/3731569.3764849},
      note = {Available: {https://arxiv.org/abs/2510.10620}}, 
}

@misc{usami2026prefilldecodeawareevaluationllminference,
  title = {Prefill/Decode-Aware Evaluation of {LLM} Inference on Emerging {AI} Accelerators},
  author = {S. Usami and V. Vishwanath and E. Wes Bethel},
  year = {2026},
  note = {Available:  {https://arxiv.org/abs/2606.17104}}
}

@misc{qwen2025qwen25technicalreport,
      title={Qwen2.5 Technical Report}, 
      author={An Yang and Baosong Yang and Beichen Zhang and Binyuan Hui and Bo Zheng and Bowen Yu and Chengyuan Li and Dayiheng Liu and Fei Huang and Haoran Wei and Huan Lin and Jian Yang and Jianhong Tu and Jianwei Zhang and Jianxin Yang and Jiaxi Yang and Jingren Zhou and Junyang Lin and Kai Dang and Keming Lu and Keqin Bao and Kexin Yang and Le Yu and Mei Li and Mingfeng Xue and Pei Zhang and Qin Zhu and Rui Men and Runji Lin and Tianhao Li and Tianyi Tang and Tingyu Xia and Xingzhang Ren and Xuancheng Ren and Yang Fan and Yang Su and Yichang Zhang and Yu Wan and Yuqiong Liu and Zeyu Cui and Zhenru Zhang and Zihan Qiu},
      year={2025},
      eprint={2412.15115},
      archivePrefix={arXiv},
      primaryClass={cs.CL},
      note = {Available: {https://arxiv.org/abs/2412.15115}}, 
}

@misc{NvdaH200,
  author       = {{NVIDIA}},
  title        = {{NVIDIA} {H200} Tensor Core {GPU}: Supercharging {AI} and {HPC} workloads},
  year         = {2026},
  howpublished = {\url{https://dam-cdn.nvd.orangelogic.com/AssetLink/5o2qgy5d2835ve2pm11i62kv8mphqta8.pdf}},
  note         = {Accessed: 2026-09-15}
}

@misc{vllm15,
  author       = {{vLLM v0.15.0}},
  title        = {{vLLM}: Easy, fast, and cheap {LLM} serving for everyone},
  year         = {2026},
  howpublished = {\url{https://docs.vllm.ai/en/v0.15.0/}},
  note         = {Accessed: 2026-09-15}
}

@misc{mirzaei2026characterizingparallelismstrategiesllm,
      title={Characterizing Parallelism Strategies in {LLM} Inference: Fundamental Compute-Communication Trade-offs}, 
      author={J. Mirzaei and J. Mitra},
      year={2026},
      eprint={2610.05305},
      archivePrefix={arXiv},
      primaryClass={cs.DC},
      note = {Available:  {https://arxiv.org/abs/2610.05305}}
}


% George Kour and Raid Saabne.
% \newblock Real-time segmentation of on-line handwritten arabic script.
% \newblock In {\em Frontiers in Handwriting Recognition (ICFHR), 2014 14th
%   International Conference on}, pages 417--422. IEEE, 2014.

% \end{thebibliography}

% \newpage
% \appendix  % Start appendix
% \section*{Appendix}

% \section{Tensor Parallelism: ITL and TTFT vs Batch Size}

\end{document}